\documentclass[
 floatfix,
 reprint,
 amsmath,amssymb,
 aps,
]{revtex4-2}

\usepackage{graphicx}% Include figure files
\usepackage{dcolumn}% Align table columns on decimal point
\usepackage{bm}% bold math
\usepackage{subfigure}
\usepackage{xcolor}
\usepackage{comment}

\begin{document}

\preprint{APS/123-QED}
\title{Detection of Topology via Entanglement Oscillations}

\author{Chunyu Tan}
 \email{chunyuta@usc.edu}
\author{Hubert Saleur}%
 \altaffiliation[Also at ]{Institut de Physique Th\'eorique, CEA Saclay, 91191 Gif Sur Yvette, France Physics.}%Lines break automatically or can be forced with \\
 \author{Stephan Haas}
\affiliation{%
 Department of Physics and Astronomy, University of Southern California, CA 90089-0484}%

\date{\today}% It is always \today, today,

\begin{abstract}
We introduce a  characterization of topological order based on bulk  oscillations of the entanglement entropy and the definition of an `entanglement gap', showing that it is  generally applicable to pure and disordered quantum  systems. Using exact diagonalization and the strong disorder renormalization group method, we demonstrate that this approach  gives results in agreement with the use of  traditional topological invariants, especially in cases where topological order is known to persist in the presence of off-diagonal bond disorder. The entanglement gap  is then used to analyze classes of quantum  systems with alternating bond types, allowing us to construct their topological phase diagrams. The validity of these phase diagrams is verified in limiting cases of dominant bond types, where the solution is known. 
\end{abstract}
%\begin{description}

%\end{description}

\maketitle

{\it Introduction - } It is well known that entanglement measures can be used to identify topologically non-trivial features in higher-dimensional many-body systems. For example, in topological (long range entangled) states of matter, the scaling of the von Neuman entropy $S(L)$ with system size has been shown to be offset by a universal constant $\gamma$, tied to topological invariants: $S(L)=\alpha L^{D-1} - \gamma + \cdot\cdot\cdot$ ~\cite{PhysRevLett.96.110404,HAMMA200522}. For particular lattice geometries,  $\gamma$ can be isolated by carefully subtracting contributions to $S(L)$ of appropriately chosen partitions of the system ~\cite{PhysRevLett.96.110405,PhysRevB.101.085136}. Furthermore, it has been shown that the entanglement spectrum, i.e. a Schmidt decomposition of a many-body quantum state, contains pertinent information about its topological properties ~\cite{PhysRevLett.101.010504,PhysRevB.80.180504}.

While  intrinsic topological order does not exist in 1D, gapped SPT (Symmetry Protected Topological) phases share many properties with their higher-dimensional, long range entangled  cousins. We show in this work how entanglement entropy can also be used in this case to detect and characterize topology. Furthermore, the new aspects we uncover can  be extended to gapless situations, as well as to higher dimensions.

A very simple example of an SPT  phase in 1D is provided by the Su-Schrieffer-Heeger model, which is in essence a chain of hopping electrons with alternating couplings ~\cite{PhysRevLett.42.1698,PhysRevB.22.2099,PhysRevB.28.1138}. The topological features  of this model have been investigated using invariants such as the winding number, the presence  of  edge states, or the  Berry phase. However, such measures are not well adapted to the introduction of interactions or disorder. Winding numbers, for instance, rely on the dispersion relation of {\sl one particle} excitations, and require subtle generalizations to be used in the context of Luttinger liquids ~\cite{PhysRevB.96.085124}. Other generalizations are necessary in the presence of disorder, since momentum is not a conserved quantum number any longer ~\cite{PhysRevLett.113.046802}. With disorder strong enough, the presence of edge states is difficult to assess in a given sample, and does not lend itself to simple `averaging'. 
Entanglement, in contrast, remains straightforward to analyze - even with interactions, since it is only a property of the ground state -  and is particularly well suited to the introduction of disorder (especially in 1D) thanks to the (real-space) strong disorder  renormalization group (SDRG) approach ~\cite{PhysRevLett.43.1434,PhysRevB.22.1305,refael2004entanglement}.

\begin{figure}[h!]
\centering
\includegraphics[scale=0.35]{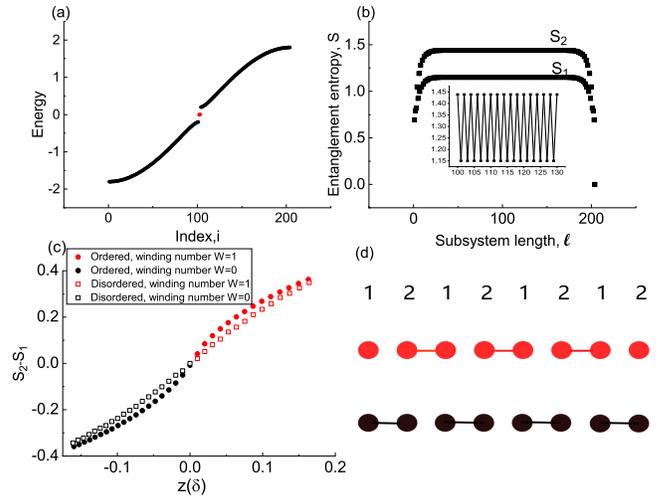}
\caption{(a) Energy spectrum of an open-ended $J_1-J_2$ SSH chain with N=2$\times$102=204 sites and couplings $J_1$=0.8 and $J_2$=1. The E=0 edge states are indicated by the red dot. (b) Dependence of the bipartite von Neumann entanglement entropy  on the  subsystem size $l$.  In the bulk ($l =O(L)$), this quantity oscillates between two values, $S_1$ and $S_2$, depending on whether the subsystem size is even or odd. The inset shows a representative segment of these bulk oscillations. The couplings  are chosen to be the same as in (a). (c) Dependence of the bulk entanglement gap, $S_2-S_1$, on the relative coupling strengths, tuning between the topological sectors with winding numbers W=1 (red) and W=0 (black). Filled symbols correspond to  the pure case, where $z=(J_2-J_1)/(J_1+J_2)$, and open symbols to the bond disordered case, where the difference of the dimerization parameter $\delta$ (defined in Eq. 3) is used as the tuning parameter instead. (d) Illustration of quantum states in the limiting cases $J_1 \rightarrow 0$ (W=1 sector with a pair of edge states) and $J_2 \rightarrow 0$ (W=0 sector with no edge states).}
\label{fig:fig1}
\end{figure}

{\it The $J_1-J_2$ model -}
To illustrate this approach, we first consider an antiferromagnetic spin-1/2  xx-chain with dimerized couplings $J_1,J_2$. This  model can be mapped via the Jordan-Wigner transformation onto the Su-Schrieffer-Heeger (SSH) model, which describes free fermions hopping on a dimerized chain.   The Hamiltonian  for an open-ended chain with an even total number of sites $N=2L$ is given by
\begin{equation}
\hat{H}=J_1\sum_{n=1}^{L} \hat{a}_{2n-1}^{\dagger} \hat{a}_{2n}+J_2\sum_{n=1}^{L-1}\hat{a}_{2n}^{\dagger} \hat{a}_{2n+1}+\mathrm{h.c.} .\label{J1J2ham}
\end{equation}

 Its energy spectrum is shown in Fig. 1(a). For $J_1\neq J_2$, there is an energy gap $\Delta=2|J_2-J_1|$, separating the bonding and antibonding bands. Furthermore, when $J_1>J_2$, the system's winding number is W=0, and there are no topological edge states, whereas when $J_1<J_2$, the winding number is W=1, and there are two exponentially localized edge states at zero energy, with a localization length $\xi=1/\ln(J_2/J_1)$ ~\cite{asboth2016short}. 

 The chain being non-interacting, all quantities of interest  can be determined from one-particle properties.  The entanglement entropy, in particular,  follows from  the correlation matrix ~\cite{peschel2009reduced}. As long as the energy gap is finite, the bipartite entanglement of contiguous spatial segments with the remainder of the system saturates in the bulk. Focusing  on such segments made of $l$ spins counted  from one end of the chain, we  observe  that the von Neumann entropy exhibits strong parity effects,  saturating for  $l =O(L)$  to an asymptotic value that depends on the parity of $l$:
\begin{equation}
\lim\limits_{l \to L}S(l)=\left\{\begin{array}{ll}{S_{1}} & {\text { if } l \text { mod } 2=1}, \\ {S_{2}} & {\text { if } l \text { mod } 2=0}.\end{array}\right. 
\end{equation}
An example of this type of oscillation of the entanglement entropy is shown in Fig. 1(b) for $J_1/J_2=0.8$. 
Furthermore, we note that parity effects for the entanglement entropy  are also known to occur  at the critical (gapless) point $J_1=J_2$ ~\cite{affleck2009entanglement}. At this point, however,  the amplitudes of the  oscillations decay as a power law with $l$, whereas in the off-critical regime {\sl bulk} entanglement oscillations persist deep inside the system. 

This leads us to introduce our 
 central observable -  the {\it bulk entanglement gap} -  defined by the magnitude of the bulk entanglement oscillations (insert of Fig. 1(b)), which in this simple model is given by $\Delta S = S_2-S_1$. As seen in Fig. 1(c), this quantity is positive in the W=1 sector, and negative for W=0, in clear correspondence with the  topological phase transition at $J_1=J_2$. 
Furthermore, defining $z=(J_2-J_1)/(J_1+J_2)$, we find that away from criticality, this quantity scales as $\Delta S \sim \ln 2 - (1-z)^2$, whereas close to criticality $\Delta S \sim -z\ln z$. Looking at other topological invariants in this model, we find that there is a finite string order parameter ~\cite{PhysRevB.40.4709} when $J_1 \neq J_2$, i.e.$z\neq 0$,  indicating broken $Z_2\times Z_2$ symmetry ~\cite{Oshikawa}. Furthermore, for open ended chains with even N we observe two edge states in the W=1 sector and no edge states for W=0. In the limit $J_1 \rightarrow 0$, these two states are depicted in Fig. 1(d), where the solid lines indicate valence bonds. Naturally, the number of dangling bonds at the edges depends on how the chain is cut. For example, for odd N, there would be one edge state in both cases. 

When there is off-diagonal disorder, $J_1$ and $J_2$ become random variables. Interestingly, we find that also in this case, the oscillations of the bulk entanglement entropy are correlated with topological invariants. 
In the bond disordered $J_1-J_2$ model, a dimerization parameter can be defined as ~\cite{FisherIsing}
\begin{equation}
\delta=\frac{\left[\ln J_{2}\right]_{\mathrm{av}}-\left[\ln J_{1}\right]_{\mathrm{av}}}{\operatorname{var}\left[\ln J_{1}\right]+\operatorname{var}\left[\ln J_{2}\right]}.
\end{equation}
Using both exact numerical diagonalization and the strong disorder renormalization group technique (SDRG), we find clear indications of a topological phase transition at $\delta = 0$. Specifically, in Fig. 1(c) $J_1$  is uniformly distributed within the interval $[0, 2\bar{J}_1]$, and $J_2$ is uniformly distributed in $[0, 2]$.
When $\delta = 0$, i.e. when  the odd  and the even bonds have the same distribution, the system is critical and in a random singlet (RS) phase. We have numerically verified that in the RS phase, the string order parameter ~\cite{PhysRevB.40.4709} vanishes, whereas when $\delta$ is not equal to 0, the system is in a random dimer (RD) phase, and hence the string order parameter is non-vanishing,  which implies that the RD phase is topologically nontrivial ~\cite{Girvin}. More specifically, we find that the string order parameter scales linearly with $\delta$ close to $\delta =0$. 

The SDRG flow equations of the $J_1-J_2$ chain are the same as for the transverse field Ising spin chain ~\cite{Girvin}. Based on previous studies, we know that when $\delta$ is small, the fixed point bond distribution is given by ~\cite{FisherIsing}
\begin{equation}
P_o(\zeta)=u_{0} e^{-\zeta u_{0}} \quad \text { and } \quad P_e(\beta)=\tau_{0} e^{-\beta \tau_{0}}
\end{equation}
with
\begin{align}
\tau_{0}&=\frac{2 \delta_{0}}{1-e^{-2 \delta_0\left(\Gamma+C_{0}\right)}} \quad \text { and } \quad u_{0}&=\frac{2 \delta_0}{e^{2 \delta_0\left(\Gamma+C_{0}\right)}-1}, 
\end{align}
where $\zeta$=$\ln(\Omega/J_o)$, $\beta$=$\ln(\Omega/J_e)$, and $\Gamma$=$\ln(\Omega_I/\Omega)$. $\Omega_I$ is the initial magnitude of the largest bond, and $\Omega$ is the corresponding magnitude  after renormalization. If the initial $\delta = \delta_0$  is small, $\delta$  remains almost a constant during the RG process. $C_0$ is an integral constant which depends on the initial bond distribution. To evaluate the entanglement entropy, we calculate the average number of singlets that form across a particular bond. For odd (even) bonds, this average number of singlets grows as $d\overline{N}_{o}=d \Gamma P_o(\zeta=0)=d \Gamma u_{0}$ ($d\overline{N}_{e}=d \Gamma P_e(\beta=0)=d \Gamma \tau_{0}$). The bulk entanglement gap can then be derived in this limit, 
\begin{equation}
\Delta S=\ln2 \times \left[\ln(e^{2\delta(\Gamma+C_{0})}-1)+\ln(1-e^{-2\delta(\Gamma+C_{0})})\right].
\end{equation}
We see that  when $\delta$ is small, $\Delta S$ scales linearly with $\delta$, but its slope is not universal, but depends on the initial distribution. Meanwhile, when $\delta\rightarrow\infty$, the system becomes a fully dimerized spin chain, and $\Delta S \rightarrow \ln 2$.

While it is difficult to discern topological edge states in the disordered case, the SDRG provides an alternative approach. When the even bonds are stronger than the odd bonds, there is a finite probability that the first spin and the last spin of the open chain form a singlet. Because a bond connecting the two edge spins can only be  formed in the last RG step, whenever such a bond occurs, the interaction between the two spins is very weak, akin to the zero energy edge states in the W=1 regime of the pure $J_1-J_2$ model. Indeed, we find 
a non-zero probability of formation of a bond connecting the two edge spins, increasing as $P \sim|\delta|^{2}$ for small $\delta$.  This non-zero probability  can be taken as an independent indication that the disordered dimerized spin-$1/2$ chain is in a topologically non-trivial sector.

\begin{figure}[h!]
\centering
\includegraphics[scale=0.35]{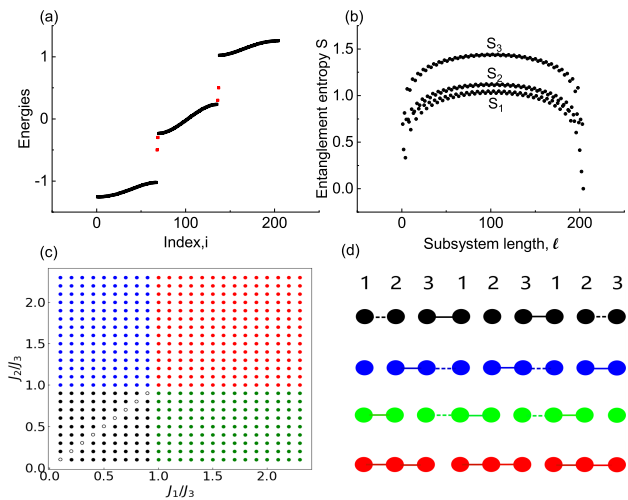}
\caption{(a) Energy spectrum of an open-ended $J_1-J_2-J_3$ antiferromagnetic xx-chain with N=3$\times$68=204 sites and couplings $J_1$=0.3, $J_2$=0.5, and $J_3$=1. The edge states are indicated by red dots. (b) Dependence of the bipartite von Neumann entanglement entropy  on the subsystem size $l$.  In the bulk, this quantity oscillates between three values, $S_1$, $S_2$, and $S_3$, depending on  the subsystem size.  Couplings  are chosen to be the same as in (a). (c) Phase diagram of the pure $J_1-J_2-J_3$ model, obtained using bulk entanglement gaps. (d) Illustration of quantum states in the limiting cases $J_1, J_2 \rightarrow 0$ (sector with two pairs of edge states), $J_2 \geq J_3 > J_1$ (sector with one pair of right edge states), $J_1 \geq J_2 > J_3$ (sector with one pair of left edge states), $J_3 \rightarrow 0$ (sector with zero edge states). Solid lines denote the strongest couplings, whereas dashed lines indicate next-strongest couplings, which give rise to pairs of finite energy edge states. }
\label{fig:fig2}
\end{figure}

{\it The $J_1-J_2-J_3$ model  - }
We now consider a model of quantum spins on a trimerized chain of length $N=3L$ and with cyclically alternating nearest neighbor couplings.
\footnote{The Hamiltonian for this model is given by 
 $\hat{H}=J_1\sum_{n=1}^{L} \hat{a}_{3n-2}^{\dagger}\hat{a}_{3n-1}
 +J_2\sum_{n=1}^{L-1}\hat{a}_{3n-1}^{\dagger} \hat{a}_{3n}
  +J_3\sum_{n=1}^{L-2}\hat{a}_{3n}^{\dagger}
 \hat{a}_{3n+1}+\mathrm{h.c.}$.}
The energy spectrum for a particular choice of parameters, $J_1$, $J_2$ and $J_3$, is shown in Fig. \ref{fig:fig2}(a). It generally consists of three bands (bonding, anti-bonding, and non-bonding) and up to two pairs of bound states (indicated in red).  
The topological features of this model are more subtle than for the SSH model with a two-site unit cell, except for $J_1=J_2$, when an inversion symmetry is present. Away from this symmetry subspace,  concepts such as mapping of the model to higher dimensions  ~\cite{PhysRevA.99.013833} and a piecewise calculation of Berry phases ~\cite{AgarwalLiu} have been invoked to motivate their topological origin. The presence of robust edge states has however been clearly identified in much of the phase diagram ~\cite{PhysRevA.99.013833, AgarwalLiu}, even away from the $J_1=J_2$ symmetry line, and is summarized in Table I. 

\begin{table}[h]
\caption{\label{tab:table1}Relationship between the oscillations of entanglement entropy, relative bond strengths, and number of edge states.}
\begin{ruledtabular}
\begin{tabular}{|c|c|p{8em}|}
\hline \text { Bulk Entanglement } & {\text { Bond Strength }} & {\text { Edge States }} \\
\hline $S_{3} > S_{2} \neq S_{1}$ & {$J_{3}>J_{2} \neq J_{1}$} & {two pairs of edge states with different energies, localized on both ends } \\
\hline $S_{3}> S_{2}=S_{1}$ & {$J_{3}>J_{2}=J_{1}$} & {two pairs of edge states with the same energies} \\
\hline $S_1 \geq S_{3}>S_{2}$ & {$J_{1} \geq J_{3}>J_{2}$} & {one pair of  edge states localized  at the left end} \\
\hline $S_2 \geq S_{3}>S_{1}$ & {$J_{2} \geq J_{3}>J_{1}$} & {one pair of edge states localized  at the right end} \\
\hline $S_1, S_{2} \geq S_{3}$ & {$J_{1}, J_{2} \geq J_{3}$} & {no edge states} \\
\hline
\end{tabular}
\end{ruledtabular}
\end{table}

Like for the two-site unit cell SSH model, we can study the entanglement entropy $S(l)$ of a region of $l$ spins counted from one end of the chain. For the $J_1-J_2-J_3$ model, we find that this entropy now exhibits oscillations between at most three different values depending on the couplings and the value of $l\hbox{ mod} 3$. An example for these oscillations is shown in Fig. \ref{fig:fig2}(b) for the same parameter set as in Fig. \ref{fig:fig2}(a). Note that, while edge states occur within gaps and the system can still be considered an insulator for appropriate fillings, our entanglement calculations are performed at half-filling (zero magnetic field in the spin-chain language), where the system is gapless. This means the curves for $S(l)$ do not saturate when $l=O(L)$, unlike in the gapped $J_1-J_2$ model. Instead, we have checked that the dependency expected from the Cardy-Calabrese formula (for open boundary conditions) holds ~\cite{Calabrese_2009}, namely
\begin{equation}
    S_l={c\over 6}\ln\left({2N\over\pi}\sin{\pi l\over N}\right)+b_{l\tiny{mod } 3}, \label{SCarCal}
\end{equation}
where the offsets can take up to three different values depending on the couplings:
the entanglement gaps are then given by  the differences of the $b_{l\tiny{mod } 3}$ in Eq.~\ref{SCarCal}.
We note that topological entropy ~\cite{PhysRevLett.96.110404,PhysRevLett.101.010504} in higher dimensions can sometimes be related with the value of the offsets $b_l$  as in (Eq. ~\ref{SCarCal}) for one dimensional projections (such as edge states in the quantum Hall case  ~\cite{Fendley}). 

The full phase diagram, shown in Fig. \ref{fig:fig2}(c),  has been obtained using the zero crossings of the corresponding entanglement gaps, analogous to Fig. \ref{fig:fig1}(c). Here we observe four distinct regimes that coincide with those derived earlier by considering edge states ~\cite{PhysRevA.99.013833, AgarwalLiu}. For $J_3 >J_1,J_2$ (black dots), there are two pairs of edge states with different energies. As mentioned above, within this regime, there is an inversion symmetric line $J_1=J_2$ where the edge state pairs become degenerate (unfilled black circles). There are also two regimes (blue and green dots) with only one pair of edge states, located either at the right or at the left end of the open chain. Finally, for $J_3<J_1,J_2$, there are no edge states at all. The corresponding ground states in the limits $J_3 \rightarrow \infty$ (black), $J_2 \rightarrow \infty$ (blue), $J_1 \rightarrow \infty$ (green), and $J_3 \rightarrow 0$ (red) are depicted in Fig. \ref{fig:fig2}(d). \footnote{In order to understand the formation of finite energy edge states in open $J_1-J_2-J_3$ chains, one needs to consider the strongest couplings, denoted by solid lines in Fig. ~\ref{fig:fig2}(d), as well as the subdominant couplings. For the configuration with two pairs of edge states (drawn in black), $J_3$ dominates, leading to the formation of a product state of alternating dimer singlets and dangling spins in the bulk. On the two edges, however, there are two pairs of spins, coupled via the subdominant $J_1$ and $J_2$ respectively, giving rise to finite energy edge bound state pairs at energies $\pm J_1$ and $\pm J_2$. Similarly, for dominant $J_2$ and subdominant $J_3>J_1$ (drawn in blue), a product state of trimers is formed in the bulk, which - together with the dangling spin on the left - combines to a gapless low-energy non-bonding band. In addition, the remaining dimer singlet on the right edge gives rise to a pair of bound states at energies $\pm J_2$.} We also note that for open chains the number of edge states depends on the position of the cuts as well as on the parameter regime of the couplings.

\begin{figure}[h]
\centering
\includegraphics[scale=0.35]{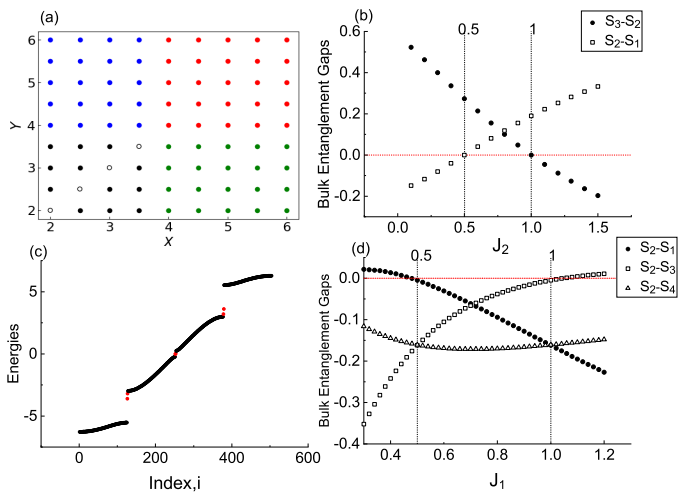}
\caption{(a) Phase diagram of the bond disordered $J_1-J_2-J_3$ model, obtained using entanglement gaps.
(b) Identification of topological phase boundaries via entanglement gaps in the pure $J_1-J_2-J_3$ model. The particular scans shown here are for fixed $J_1=0.5$ and $J_3=1$, identifying transitions between to 2 edge states regimes at $(J_2)_{c1}=0.1$ and these 2 edge states have the same energy $(J_2)_{c2}=0.5$.
(c) Energy spectrum of an open-ended $J_1-J_2-J_3-J_4$ antiferromagnetic xx-chain with N=4$\times$126=504 sites and couplings $J_1$=3, $J_2$=2,  $J_3$=2.5, and $J_4$=4.5. The topological edge states are indicated by red dots. 
(d) Identification of topological phase boundaries via entanglement gaps in the pure $J_1-J_2-J_3-J_4$ model. The  scans shown here are for fixed $J_2=0.5$, $J_3=0.5/J_1$, $J_4=1.0$, identifying phase transitions from one pair of edge states to two pairs of edge states at $(J_1)_{c1}=0.5$ and from two pairs of edge states to one pair of edge states at $(J_1)_{c2}=1.0$. 
}
\label{fig:fig3}
\end{figure}

As mentioned above, one major distinguishing feature of the entanglement gap measure compared to other topological invariants is that it can be easily applied to disordered systems. As an illustrative example, in Fig. \ref{fig:fig3}(a) $J_3$ is uniformly distributed within the interval [3,5], $J_1 \in [x-1,x+1]$,  and $J_2 \in [y-1,y+1]$. 
There are $N= 3\times 68 = 204$ sites and, we consider  200 realizations. Interestingly, the topology of the phase diagram of this disordered system closely resembles that of the pure system in Fig. \ref{fig:fig2}(c), featuring four regimes with distinct numbers of edge states. 

Fig. \ref{fig:fig3}(b) illustrates how the phase boundaries in Figs. \ref{fig:fig2}(c) and \ref{fig:fig3}(a) are drawn, i.e. via zero crossings of the corresponding entanglement gaps. Specifically, in Fig. \ref{fig:fig3}(b), we show a cut through the phase diagram of the pure $J_1-J_2-J_3$ chain, fixing $J_1=0.5$, $J_3=1$, and scanning $J_2\in [0,1.5]$, focusing on the bulk entanglement gaps $S_3-S_2$ and $S_2-S_1$: when  $S_3-S_2$=0, there is a transition from a regime with two pairs of edge states to a regime with one pair of edge states, and at  $S_2-S_1$=0, these 2 edge state pairs have the same energy (inversion symmetric point). 

{ \it  The $J_1-J_2-J_3-J_4$ model - }
As a final  example, let us consider the open $J_1-J_2-J_3-J_4$ chain ~\cite{Maffei_2018,xie}, whose energy spectrum is shown in Fig. \ref{fig:fig3}(c) for an $N=4\times 126=504$ system and a particular coupling choice at which we observe multiple edge modes.\footnote{The Hamiltonian for this model is given by 
 $\hat{H}=J_1\sum_{n=1}^{L} \hat{a}_{4n-3}^{\dagger}\hat{a}_{4n-2}
 +J_2\sum_{n=1}^{L-1}\hat{a}_{4n-2}^{\dagger} \hat{a}_{4n-1}
  +J_3\sum_{n=1}^{L-2}\hat{a}_{4n-1}^{\dagger} \hat{a}_{4n}
    +J_4\sum_{n=1}^{L-3}\hat{a}_{4n}^{\dagger} \hat{a}_{4n+1}
  +\mathrm{h.c.}$.} This model generally has a central gap, and  winding numbers can be straightforwardly defined, with the exception of degeneracy subspaces, indicating topological criticality. Fig. \ref{fig:fig3}(d) displays some of the pertinent entanglement gaps along a scan through the model's phase space. For this particular example, we set $J_2=0.5, J_3=0.5/J_1, J_4=1$, while tuning $J_1 \in [0.3,1.2]$. Here we observe that when  $S_2-S_1 = 0$ ($J_1 =  0.5$), there is a  transition from a regime with one pair of edge states to a regime with two pairs of edge states, and when $S_2-S_3  = 0$ ($J_1  = 1$), there is a second transition from a regime with two pairs of edge states to a regime with one pair of edge states. 

{\it Conclusions - }
The use of entanglement gaps to identify and characterize topological phases should not be restricted to one-dimensional systems. For example, this measure lends itself quite naturally to higher dimensional valence bond solids, and more generally to tensor product states. However, let us also point out some limitations of this procedure. Entanglement measures  naturally depend on the choice of tensor decomposition of the Hilbert space ~\cite{Caban_2005}, and in some cases it is not straightforward to find an appropriate representation. For example, integer spin Heisenberg chains are known to have topologically non-trivial ground states, but one would likely need to resort to an AKLT type valence bond representation in order to observe entanglement gaps ~\cite{PhysRevLett.59.799}. 
In conclusion, it is fair to ask whether the presence of an entanglement gap results ultimately from some sort of (imposed or spontaneous) dimerization, or whether it could be construed as some variant of topological entropy - or remnant thereof in 1D. Therefore it will be important to study higher dimensional topological models with the same approach.

{\it Acknowledgements - }
We would like to thank Lorenzo Campos Venuti, Anuradha Jagannathan, Yu Yao, Zhihao Jiang and Henning Schl\"omer for useful discussions. 
This work was supported by the US Department of Energy under grant number DE-FG03-01ER45908. 

\bibliography{references}
\end{document}